\documentstyle[11pt,paspconf]{article}
\begin{document}

\title{A Numerical Experiment of a Triple Merger of Spirals}

\author{H\'ector Aceves}
\affil{IAA-CSIC. Apdo. Postal 3004. Granada 18080, Spain. $\;${\sf aceves@iaa.es}}


\section{Introduction}

Several studies have addressed the topic of spiral interactions in pairs or in a small group (Barnes 1998). However there is no specific study, to our knowledge, of interacting triplets of spirals.  Here we present a preliminary report of an ongoing research on the dynamics of triplets that includes galaxies of different morphology. In particular we follow  a triplet of equal-mass spirals up to a merger remnant that resembles an elliptical-like galaxy has been formed.  

It is known that the HI component of spirals is a good tracer of interactions. Indeed, several triplets (e.g. in M81 group) that do not appear to be perturbed in optical images do so in HI maps. Although a self-consistent treatment of gas requires the inclusion of e.g. pressure forces, we use here a test-particle approach to study it during a triple interaction of spirals; this can yield information about the large-scale kinematical features of gaseous tails.

\section{The Numerical Experiment}

	We use Galaxy-like spirals with total mass $M\approx 5.8\times 10^{11}$M$_\odot$, $R_{\rm halo}\approx 135$ kpc, and disc mass $M_d \approx 4.4\times 10^{10}$ M$_\odot$ (Kuijken \& Dubinski 1995, Model B). For computational economy, the number of particles used for each spiral were: $N_{\rm bulge}=1000$, $N_{\rm disc}=5000$ and $N_{\rm halo}=12000$.  In isolation the spiral was relatively stable for a few orbital periods. Discs' orientations were obtained from three random sets of Euler angles (some animated gifs of the simulations are at {\tt http://www.iaa.es/ae/triplets.html}).

	The initial positions and bulk velocities of galaxies were obtained from one cold-collapse simulation of spherical galaxy triplets, in which a relatively rapid triple merger was found (Aceves 1999). We started the simulation at $t_0 \approx 2$ Gyr from turn-around, when two of the galaxies G1 \& G2 had their halos just overlapping.  These galaxies  were approaching at a relative velocity of $\sim 140$ km/s; the third galaxy's (G3) halo had not yet `touch' the binary's common halo and was infalling at $\sim 40$ km/s. In the following, times quoted will be the elapsed time since this $t_0$. The simulation lasted for another $\approx 12$ Gyr.

	The `gas' particles were distributed uniformly in the plane of the discs extending twice the stellar component. Forces on them were calculated using the mass interior to their position, with respect to each galaxy centre.
 	The distribution of mass in galaxies was assumed  to remain the same throughout the simulation and that it had a radial dependence.  We consider this a somewhat better approximation than assuming that all the mass of a galaxy resides at its centre; no important extra computational effort was required in this approach.

\section{Results}

\begin{figure}
\centering
\plotfiddle{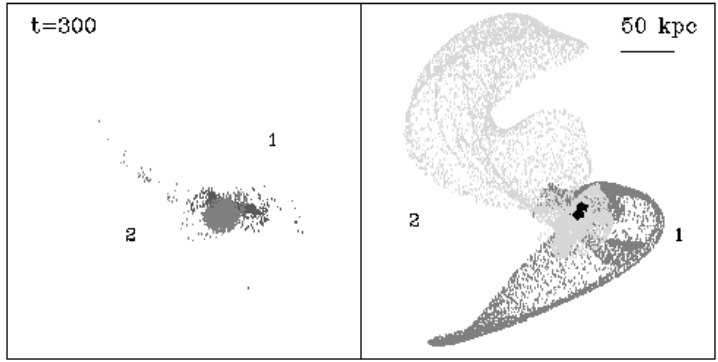}{4.5cm}{0}{55}{55}{-125}{-5}
\caption{({\it left}) $XY$-projection of disc particles of galaxies G1 \& G2 at $\approx 4$ Gyr from $t_0$. ({\it right}) Gas particles of galaxies, as well as the bulges, are shown. Numbers are intended to follow to each galaxy.}
\label{fig:t300}
\end{figure}

\begin{figure}
\centering
\plotfiddle{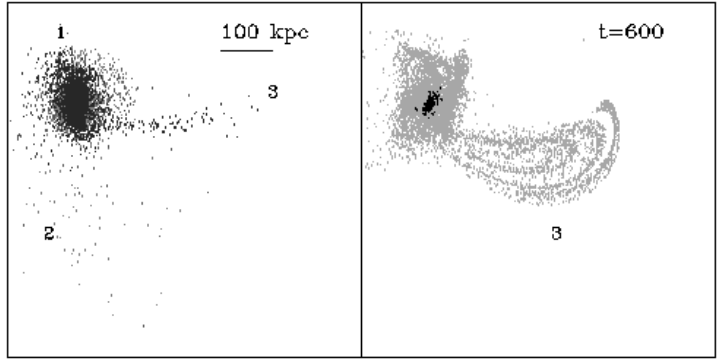}{4.5cm}{0}{55}{55}{-125}{-5}
\caption{Similar as in Fig.~\ref{fig:t300}, but at $\approx 10$ Gyr.  Note the change of scale. The `gas' tail is just separating from the stellar one in G3, as  seen in some observations. The bulge of G3 is shown on the right.}
\label{fig:t600}
\end{figure}

In Fig.~\ref{fig:t300} we show a projection of galaxies G1 \& G2 discs interacting at $\approx 4$ Gyr; galaxy G3 has not yet come close enough for its halo to be perturbed. The discs show large damages and a long stellar tidal tail has developed in G2 ($\sim 150$ kpc, in projection). The gaseous tails generated reach $\sim 200$ kpc in projection.  Some stars and gas particles are {\it returning} to the `binary' galaxy due to conservation of angular momentum and because they do not have enough kinetic energy to leave the local potential well of the binary.
 	At $\approx 10$ Gyr galaxies G1 \& G2 have already merged with G3; see Fig.~\ref{fig:t600}.  Galaxy G3 shows a very large projected stellar tidal tail of $\sim 350$ kpc, which is about the same size as its gaseous one.  At this time the tidal tail of G2 has almost disappear, although important traces of it exist as a more diffuse component. 

\begin{figure}
\centering
\plotfiddle{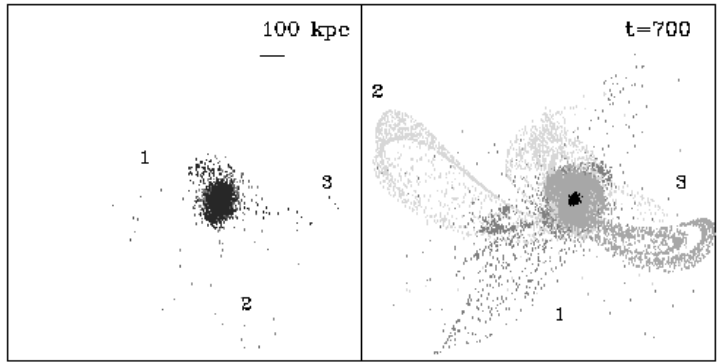}{4.5cm}{0}{55}{55}{-125}{-5}
\caption{Final configuration. Note again the change in scale, and how some stars are very far away from the central region. }
\label{fig:merger}
\end{figure}

\begin{figure}
\centering
\plotfiddle{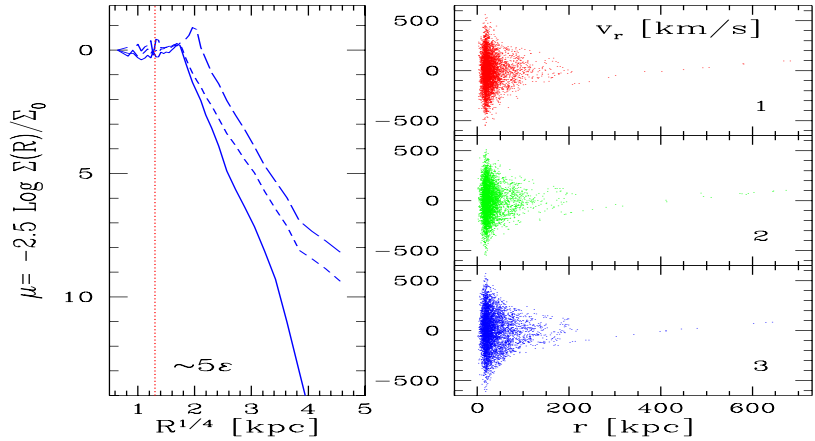}{4.7cm}{0}{55}{30}{-165}{-50}
\caption{({\it left}) Three orthogonal surface number density profiles of luminous matter at $t\approx 12$ Gyr. ({\it right}) Phase-space portraits of discs. Note that signatures of the encounter in the three galaxies persist.}
\label{fig:fase}
\end{figure}

  The configuration at the end of the simulation is shown in Fig.~\ref{fig:merger}.  Enormous gaseous tails $\sim 1$ Mpc are present and the stellar tidal tail of G3 seen in Fig.~\ref{fig:t600} is still present. Since the gas thermodynamics has not been considered, it is probable that some `HI gas' has been ionized by shocks developed during the interaction and, hence,  this extension can be an upper limit; also, the initial size of the gaseous discs and the spiral matter profile determine this extent.
	After removing all unbound particles the shape of the merger was found to be prolate. No disc or bulge particles were formally found to be unbound, but galaxies G1 \& G2 had lost $\approx 3$\% of their halo mass and G3 $\approx 10$\%. The merger remnant shows a luminous profile that resembles that of an elliptical, although with a rather isothermal core, its phase-space has retained more information about its spiral past; see Fig.~\ref{fig:fase}. The physical size of the stellar tails reach $\sim 500$ kpc!

\begin{figure}
\centering
\plotfiddle{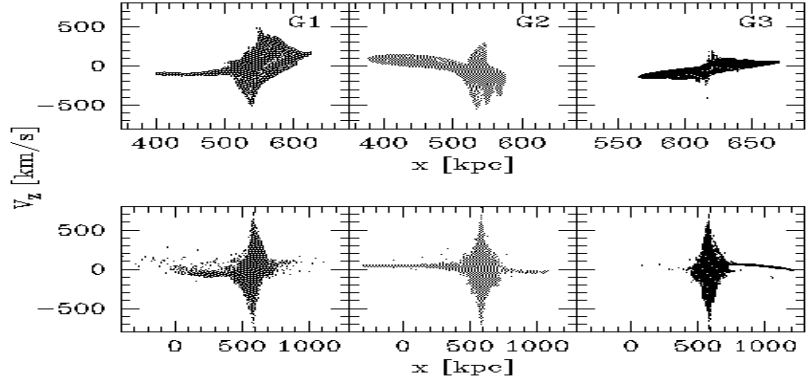}{4.5cm}{0}{60}{36}{-155}{-11}
\caption{Velocity profiles of gas along the $X$-axis at $\approx 4$ Gyr 
({\it top}), and at $\approx 12$ Gyr  ({\it bottom}), for each spiral galaxy.}
\label{fig:gas}
\end{figure}

	Figure~\ref{fig:gas} shows velocity profiles for the gas, which would mimic the result of placing a narrow slit along the $X$-axis, both at $\approx 4$ Gyr ({\it top}) and 12 Gyr ({\it bottom}). Velocities along the {\it l.o.s.} of $\sim 100$ km/s in the outer parts with a width of $\sim (50,100)$ km/s are obtained, respectively, from G1 \& G2 at $\approx 4$ Gyr. At the end of the simulation G1 shows a broad dispersion in the velocities, while galaxies G2 \& G3 show a more well defined width of $\sim 50$ km/s in the $XV_z$-plane. Tidal tails in G3 developed very late in the merging process.

\section{Final Comments}

Some particular results suggested by the present simulation, with its specific cold-collapse initial conditions, tend to show that:

\begin{itemize}
\item[{\it i.}] Encounters of spirals can effectively destroy stellar discs in $\la t_{\rm Hubble}$. Large projected stellar tidal tails $\sim 200$ kpc can develop during a binary formation, and up to $\sim 350$ kpc in the 3rd. when this joins in. However tidal tails are not eternal, they can fall-back to the remnant of their `parent' galaxies in $\la t_{\rm Hubble}$, although important traces of them remain for $\ga t_{\rm Hubble}$.

\item[{\it ii.}] Diffuse light, due to stars stripped from the discs, is to be expected in deep imaging of spiral interactions and mergers. However if this stripping occurred more than a stellar evolution time-scale ago one may not detected it, but stellar remnants  may be in the {\it field}.

\item[{\it iii.}] Gaseous tidal tails may extend to $\sim 1$ Mpc in radius, similar to the extent of the common dark halo formed, in a triple interaction in $\sim t_{\rm Hubble}$. This result suggests that gaseous halos may trace the size of a common dark halo developed during galaxy interactions. 

\item[{\it iv.}]  Although the stellar merger remnant shows an elliptical-like profile, it is more physically justifiable to look for merger signatures in velocity-space; albeit the observational difficulties. 
\end{itemize}

\acknowledgments
{\normalsize The author thanks the Spanish Government (MUTIS Program) and J. Perea  (DGICYT Project PB96-0921) for financial support.}

\end{document}